# Observation of second-harmonic generation induced by pure spin currents

Lalani K. Werake & Hui Zhao

*Department of Physics and Astronomy, The University of Kansas, Lawrence, Kansas 66045, USA*

**Extensive efforts are currently being devoted to developing a new electronic technology, called spintronics, where the spin of electrons is explored to carry information[1,2]. Several techniques have been developed to generate pure spin currents in many materials and structures[3–10]. However, there is still no method available that can be used to directly detect pure spin currents, which carry no net charge current and no net magnetization. Currently, studies of pure spin currents rely on measuring the induced spin accumulation with optical techniques[5,11–13] or spin-valve configurations[14–17]. However, the spin accumulation does not directly reflect the spatial distribution or temporal dynamics of the pure spin current, and therefore cannot monitor the pure spin current in a real-time and real-space fashion. This imposes severe constraints on research in this field. Here we demonstrate a second-order nonlinear optical effect of the pure spin current. We show that such a nonlinear optical effect, which has never been explored before, can be used for the non-invasive, non-destructive, and real-time imaging of pure spin currents. Since this detection scheme does not rely on optical resonances, it can be generally applied in a wide range of materials with different electronic bandstructures. Furthermore, the control of nonlinear optical properties of materials with pure spin currents may have potential applications in photonics integrated with spintronics.**

Our experiments are motivated by a recent theoretical prediction of second-order nonlinear optical effects induced by pure spin currents[18]. Such an effect originates from a subtle imbalance of the Faraday rotation of electrons with opposite spin orientations. An electron with a certain spin orientation causes Faraday rotation of a linearly polarized light, with an angle determined by the detuning between the frequencies of the light and the interband transition of the electron.[5,11] In a pure spin current, each electron is accompanied by another electron with an opposite crystal momentum and an opposite spin orientation. The Faraday rotation caused by the two electrons seems to cancel. However, it has been discovered that if the electrons are driven by an optical field, the work done by the intraband acceleration leads to opposite renormalizations to the interband transition frequencies at opposite momenta, since at any particular instant of time, one electron accelerates while the other decelerates.[18] Thus, the Faraday rotation caused by the two electrons is not exactly canceled, leaving a net second-order nonlinear optical susceptibility.[18]

We use a quantum interference and control technique[19] to instantaneously inject a pure spin current in a sample with a well-controlled spatial distribution. Specifically, a 400-nm-thick GaAs sample cooled to 10 K is simultaneously illuminated with two laser pulses (see Supplementary Information for details). Electrons can be excited from the valence band to the conduction band by two-photon absorption (red vertical arrows in Fig. 1b) of an $\hat{x}$-polarized, 75-fs pulse with a central wavelength of 1500 nm (red waves in Fig. 1) or one-photon absorption (green vertical arrow in Fig. 1b) of a $\hat{y}$-polarized, 290-fs pulse with a central wavelength of 750 nm (green waves in Fig. 1). Both pulses are incident along (001) direction of the sample (defined as $\hat{z}$) and are tightly focused to 2-3 $\mu$m (full width at half maximum). Due to the interference of the two transition pathways, electrons with opposite spin orientations along $\hat{z}$ (orange and blue spheres in Fig. 1) are injected with opposite average velocities along $\hat{x}$. The average velocity, $v$, and therefore the charge current density of each spin system, $J$, is proportional to $\cos(\Delta\phi)$, where $\Delta\phi$ is the relative phase between the two transition amplitudes[19]. Since the charge currents carried by the two spin systems are equal in magnitude but opposite in sign, there is no net charge current along $\hat{x}$. By analyzing the movement of the electrons with a high-resolution pump-probe technique[13], we roughly estimate that $v$ is on the order of 30 nm/ps, much higher than a typical drift velocity under an electric field. Hence, even with a moderate carrier density of $1.2 \times 10^{18} \text{cm}^{-3}$, a very high peak current density on the order of $10^5 \text{A cm}^{-2}$ is achieved.

We demonstrate the second-order nonlinear optical effects of the injected pure spin current by observing a second-harmonic (SH) generation process. An 170-fs probe pulse with a central wavelength of 1760 nm and a pulse energy of 0.1 nJ is incident on the sample along $-\hat{z}$ and is tightly focused to a spot size of 2.1 $\mu$m (black waves in Fig. 1). Although the probe pulse propagates along a direction on which the GaAs crystal is centrosymmetric, the inversion symmetry is broken by the pure spin current, allowing second-order optical responses. The SH pulse induced by the pure spin current has a central wavelength of 880 nm (purple waves in Fig. 1), with an amplitude $E_\text{J} \propto \chi_\text{J}^{(2)} E_\text{p}^2$, where $E_\text{p}$ is the field amplitude of the probe pulse and $\chi_\text{J}^{(2)}$ is the second-order nonlinear susceptibility induced by the spin current. The probe pulse is linearly polarized along $\hat{x}$, since the most efficient SH generation occurs when the polarization of the probe light is parallel to the spin current propagation direction.[18]

We detect the expected SH pulse with a coherent detection scheme, where the weak SH signal is amplified by a vectorial addition with a so-called local oscillator[20]. The local oscillator has the same frequency and ideally the same phase of the signal, but with a much stronger amplitude $E_{\text{LO}}$. In our experiments, the SH generated at the sample surface[21] is used as a natural local oscillator for simplicity. The total SH intensity is a result of the interference of these two fields: $I = (c\epsilon_0/2)(E_{\text{LO}} + E_{\text{J}})^2$, where $c$ and $\epsilon_0$ are the speed of light and the dielectric constant in a vacuum, respectively. We write this total SH intensity as $I_{\text{LO}} + \Delta I$, where $I_{\text{LO}} = (c\epsilon_0/2)E_{\text{LO}}^2$ is the intensity of the local oscillator, and $\Delta I = (c\epsilon_0/2)(2E_{\text{LO}}E_{\text{J}} + E_{\text{J}}^2)$ is the change of the total intensity due to the pure spin current.

The SH signal is sent to a silicon photodiode, and the intensity is integrated in both time and space to obtain the average power of the beam. The power corresponding to $I_{\text{LO}}$, $P_{\text{LO}}$, is measured by modulating the intensity of the probe pulse with an optical chopper, with the current-injecting pulses blocked. We find that $P_{\text{LO}}$ = 4 nW. The optical power corresponding to $\Delta I$, $\Delta P$, is measured by modulating the average velocity of each spin system, and therefore the current density, with an electro-optic phase modulator[22] (see Supplementary Information for details). Under our experimental conditions, the maximum value of $\Delta P$ is about 200 times lower than $P_{\text{LO}}$ (see Fig. 2 below), indicating that $E_{\text{LO}} \gg E_{\text{J}}$. Therefore, the second term in $\Delta I$ is negligible, and $\Delta P \propto \chi_{\text{J}}^{(2)}$.

We measure the $\Delta P$ as we systematically vary three controllable parameters in our experiments: the time delay between the peaks of the probe and the current-injecting pulses, $\tau$; the distance between the centres of the probe and the current-injecting spots, $x$; and the relative phase $\Delta\phi$.

First, Fig. 2a shows how $\Delta P$ varies with $\tau$ and $\Delta\phi$, with a fixed $x = 0$. At each $\tau$, $\Delta P \propto \cos(\Delta\phi)$ (Fig. 2b). Since $J \propto \cos(\Delta\phi)$, the linear relation between $\Delta P$ and $J$, and therefore between $\chi_J^{(2)}$ and $J$, is confirmed[18]. With a certain $\Delta\phi$, $\Delta P$ increases to a peak at about -0.06 ps, and then decays rapidly (Fig. 2c). This temporal behaviour indicates that the relaxation time of the spin current is shorter than the laser pulses at such a high carrier density. Figure 2d shows how $\Delta P$ varies with $x$ and $\Delta\phi$, with a fixed $\tau$ = -0.06 ps. The same cosine dependence on $\Delta\phi$ is observed at every probe position. At each $\Delta\phi$, $\Delta P$ has a Gaussian-like spatial profile (Fig. 2e), consistent with the size and the shape of the laser spots.

The measurements described above are repeated with different carrier densities by adjusting the power of the current-injecting pulses. In this way the injected current density is varied by changing the carrier density, while the average velocity is kept unchanged. A few examples are shown in Fig. 3. As we increase the carrier density, the peak shifts to earlier times, and the relaxation after the peak becomes faster. Both features are consistent with a faster relaxation of the spin current due to an increased carrier scattering rate. Furthermore, as summarized in the inset of Fig. 3, the height of the peak increases with the carrier density. The slight deviation from a linear relation can be attributed to the fact that, although the injected current density is proportional to the carrier density, the current relaxes faster with higher densities, and the height of the peak is determined by both the injection and the relaxation processes.

We have shown in Figs. 2 and 3 that the dependence of the observed SH signal on the probe delay, the probe position, the average velocity, and the carrier density are all consistent with a pure spin current-induced SH generation process. We have also investigated other possible effects of the current-injecting pulses. First, we can exclude any direct interaction between the laser pulses that does not involve carriers, since the temporal shape of the resulting SH would be independent of the carrier density. This is not what we observed in Fig. 3. The carrier-related effects of the current-injecting pulses include injections of a carrier density, a spin density, and a transverse charge current. We have investigated SH generation due to each of these on the same sample and under the same conditions, and found that each of them would give a much smaller signal with significantly different temporal behaviours (see Supplementary Information for details).

Accurate determination of the magnitude of this nonlinear effect is difficult because the current density is not precisely known. As a rough estimation, we assume a perfect phase match in the SH generation, and solve coupled-wave equations[21]. Such a simplification is justified since the sample thickness is smaller than the coherence length. By using the measured $P_{\text{LO}}$ and $\Delta P$, we estimate the magnitude of $\chi_{\text{J}}^{(2)}$ to be on the order of $10^{-13}$ to $10^{-14}$ m/V with $J = 10^5 \text{A cm}^{-2}$. The direction of the pure spin current can be determined by measuring the dependence of the SH signal on the polarization of the probe pulse, since the maximum SH signal is expected when the probe pulse is polarized along the direction of the current.[18] The sign of the pure spin current is related to the sign of $\chi_{\text{J}}^{(2)}$, which can be determined in a coherent detection scheme with a phase-controlled local oscillator.

The demonstrated second-order nonlinear optical effect is large enough to detect low density spin currents. We choose the probe photon energy to be less than half of the energy bandgap of the sample, so that neither the two-photon absorption of the probe pulse nor the one-photon absorption of the SH pulse is allowed. Therefore, the probe pulse will not disturb the spin current by injecting additional carriers. However, one could tune the probe photon energy toward the interband transition frequency of the electrons to increase $\chi_J^{(2)}$.[18] In our experiments, a high-repetition-rate (82 MHz) laser system is used. With high density spin currents on the order of $10^5 {\rm A \ cm}^{-2}$, the probe beam with an average power of 10 mW generates SH signals on the order of 10 pW. Due to the quadratic dependence of the SH field amplitude to the probe filed amplitude, the SH generation process can be significantly enhanced with amplified laser systems with lower repetition rates. For example, a 1-KHz laser system with the same average power (commercially available) will increase the SH power by about five orders of magnitude. Therefore, a typical pure spin current of 1 A cm$^{-2}$ generated by, for example, the spin Hall effect[5], will induce the same 10-pW SH signal. The detection can also be significantly improved by replacing the silicon photodiode with a photodetector with femtowatt detectability, which are widely available. Furthermore, the coherent nature of the SH pulse allows amplification of the signal by using a strong local oscillator in the coherent detection. In our experiments, the SH generated at the sample surface is used as the local oscillator for simplicity. When necessary, the signal can be further amplified by using an externally generated strong SH pulse.

In conclusion, we have demonstrated second-harmonic generation induced by pure spin currents, and used it for the fast, non-invasive, and non-destructive imaging of pure spin currents. Although optical interband absorptions are used to *generate* the pure spin currents for our demonstration experiment, the *detection* scheme does not rely on optical resonances. Therefore, it can be generally applied to a wide range of materials with indirect bandgaps or with bandgaps that are too small or too large for absorption-based optical detection techniques. Finally, since a pure spin current is composed of two spin-polarized charge currents with opposite spin polarizations and opposite propagation directions, each component contributes to half of the nonlinear susceptibility[18]. Therefore, although we detected pure spin currents in our experiments, this technique can also be used to study spin-polarized charge currents.

**Supplementary Information**   is linked to the online version of the paper at www.nature.com/nature.

**Acknowledgement**   We thank J. Wang, B-F. Zhu, and R-B. Liu for sharing their results prior to publication and acknowledge R-B. Liu for discussions on the experimental approaches and explaining to us the physics mechanisms of the effect. We thank John Prineas for providing us with high quality GaAs samples. This material is based upon work supported by the National Science Foundation of USA under Grant No. DMR-0954486.

**Author Contributions**   LKW constructed the experimental apparatus and performed the measurements; HZ proposed the topic, provided guidance on the experiments, and prepared the manuscript. Both contributed in data analysis and interpretations.

**Competing Interests** The authors declare that they have no competing financial interests.

**Correspondence** Correspondence and requests for materials should be addressed to H.Z. (email: huizhao@ku.edu).

**Figure 1** Schematics of the experimental configuration to observed the second-harmonic generation induced by pure spin currents. Panels a and b show the configuration in real space and energy space, respectively. The GaAs sample is simultaneously illuminated by two laser pulses (red and green waves). Quantum interference between the transition pathways driven by the two pulses (vertical red and green arrows in b) causes electrons with opposite spin orientations to be excited to energy states with opposite momenta (orange and blue spheres). Since the two spin systems move along opposite directions, a pure spin current is formed. The nonlinear optical effect of the injected pure spin current is studied by detecting second-harmonic generation ($E_\mathrm{J}$) from a probe pulse ($E_\mathrm{P}$).

**Figure 2** Second-harmonic generation induced by the pure spin current. a: The $\Delta P$ measured as functions of the probe delay, $\tau$, and the relative phase, $\Delta\phi$, with the probe and the current-injection spots overlapped ($x$ = 0). b and c: Cross sections of a, with $\tau = -0.06$ ps and $\Delta\phi$ = 0, respectively. d: The $\Delta P$ measured as functions of $x$ and $\Delta\phi$, with a fixed $\tau$ = -0.06 ps. e: A cross section of d, with $\Delta\phi$ = 0.

**Figure 3** Time evolutions of $\Delta P$ at various carrier densities. The $\Delta P$ is measured with $x = 0$ and $\Delta\phi = 0$. The carrier densities are [from bottom (black) to top (green)] 3.6, 4.8, 6.0, 7.2, 9.6, and $12 \times 10^{17} \mathrm{cm}^{-3}$, respectively. The inset shows the height of the peak as a function of carrier density.

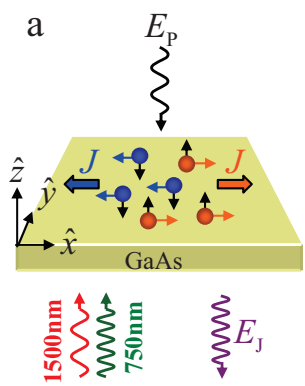 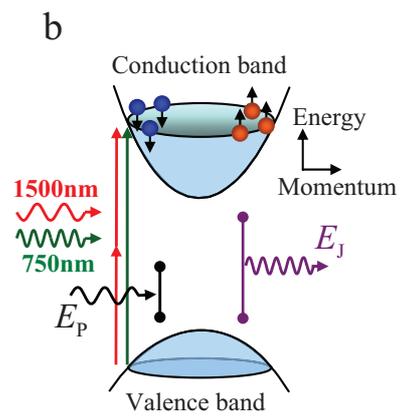

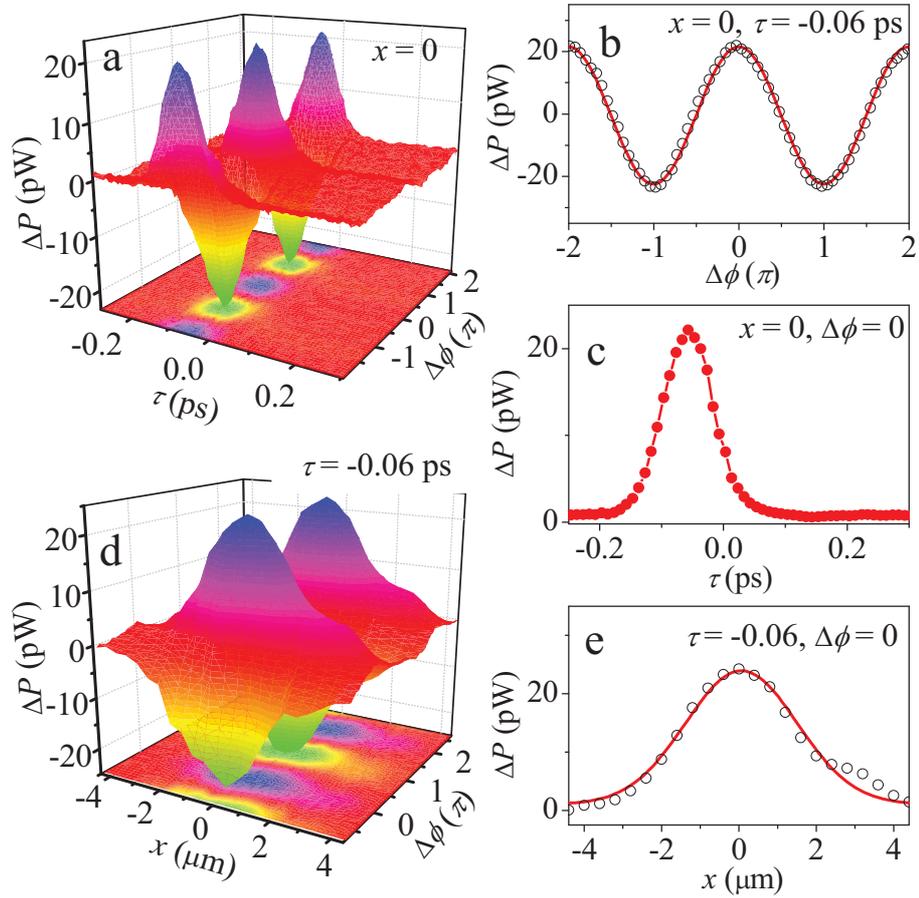

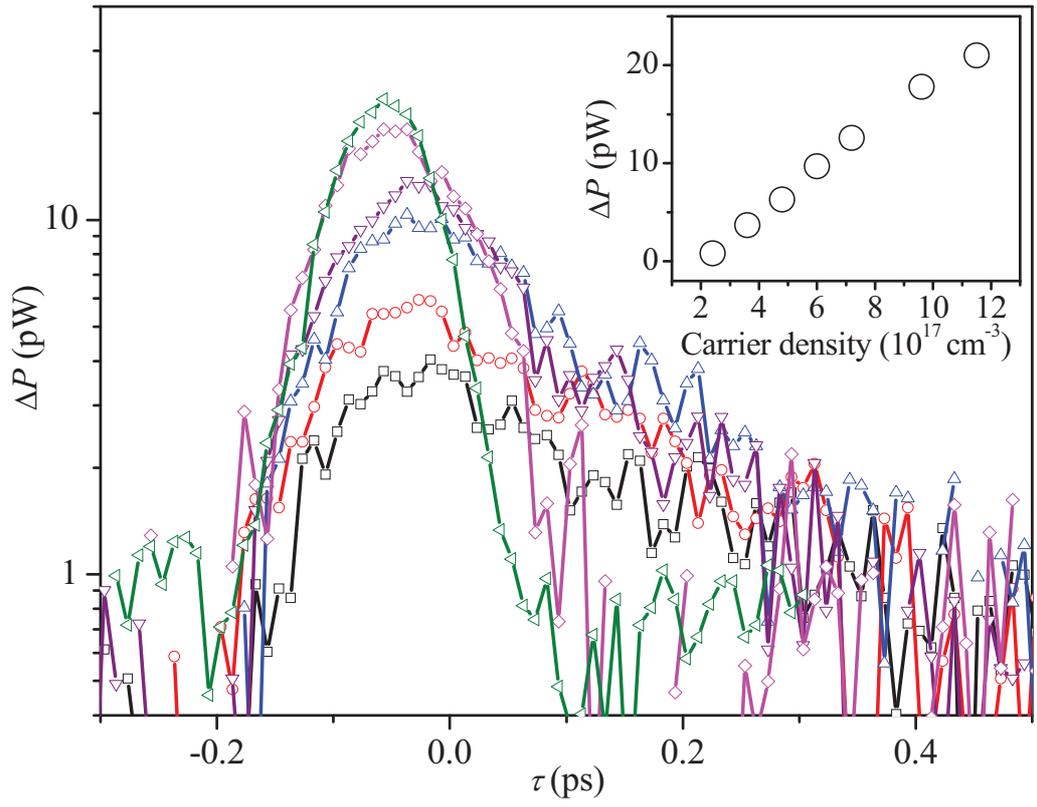

# Supplementary Discussion

## 1 Experimental setup

Figure s1 shows schematically the experimental setup. The 82-MHz ultrafast laser system (Spectra-Physics) is composed of a diode-pumped solid state laser (Millennia), a Ti:sapphire laser (Tsunami), and an optical parametric oscillator (Opal). One current-injecting pulses with a central wavelength of 1500 nm (red lines in Fig. s1) is obtained directly from the signal output of the Opal. The other current-injecting pulse with a central wavelength of 750 nm (green lines in Fig. s1) is obtained by second-harmonic generation from the 1500-nm pulse with a beta barium borate (BBO) crystal. The two pulses are separated by a dichroic beamsplitter. The 750-nm pulse is sent through an electro-optic crystal in order to modulate its phase. A Treacy grating pair is used to partially compensate for temporal broadening of the 750-nm pulse caused by the electro-optic crystal and other optics. The phase of the 1500-nm pulse is fine controlled by a retroreflector attached to a piezoelectric transducer (not shown in Fig. s1).

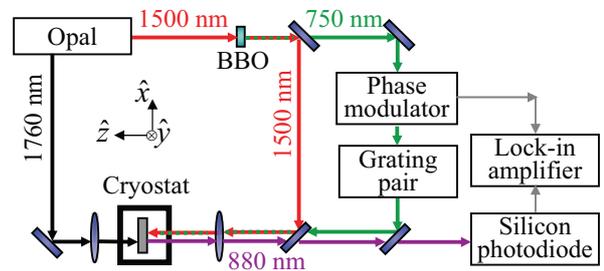

Figure s1: Schematics of the experimental setup.

The two pulses are combined by using another dichroic beamsplitter, and then focused to the GaAs sample by a microscope objective lens with a numerical aperture of 0.26. The sample is kept at 10 K in a closed-cycle cryostat (Advanced Research Systems). The 750-nm pulse has a spot size of 1.9 $\mu$m (full width at half maximum), injecting carriers with a density profile of the same size through one-photon absorption. Since the 1500-nm pulse injects carriers through two-photon absorption, the carrier density profile produced is a factor of $\sqrt{2}$ narrower than the laser intensity profile. Therefore, we set the spot size of the 1500-nm pulse to 2.6 $\mu$m, such that the two pulses produce the carrier density profiles of the same size. The temporal width of the 1500-nm pulse is the transformation-limited 75 fs, while the 750-nm pulse has a temporal width of 290 fs, due to dispersion of the electro-optic crystal and other optics that cannot be fully compensated for by the grating pair. However, such a long 750-nm pulse doesn't significantly influence the temporal resolution. Since the currents are injected through the quantum interference of the two transition pathways driven by the two pulses, the resolution is mainly determined by the shorter pulse. In each experiment, the powers of the two pulses are controlled so that each pulse injects half of the total carrier density. The polarization of each pulse is controlled by a series of wave-plates and polarizers. To inject a pure spin current with velocities along $\hat{x}$ and spin orientations along $\hat{z}$, the 1500-nm and 750-nm pulses are linearly polarized along $\hat{x}$ and $\hat{y}$, respectively[19].

The probe pulse with a central wavelength of 1760 nm (black lines in Fig. s1) is obtained from the idler output of the Opal, and is focused to the sample by another objective lens with a numerical aperture of 0.40. The spot size and the temporal width at the sample are 2.1 $\mu$m and 170 fs, respectively. The generated second-harmonic (SH) pulse with a central wavelength of 880 nm (purple lines in Fig. s1) is collimated and detected by a silicon photodiode. A combination of colour filters are used in front of the photodiode in order to block all unwanted light, which includes light from the transmitted probe pulse, the reflected current-injecting pulses, and also the photoluminescence from the GaAs sample.

The voltage signal from the photodiode is measured with a lock-in amplifier that is referenced to the phase-modulation frequency. As described in the main text, the quantum interference and control technique allows us to control the magnitude of the current density by changing the relative phase between the two transition amplitudes, $\Delta\phi$. The average velocity of each spin system is proportional to $\cos(\Delta\phi)$. By applying a square-wave voltage signal to the electro-optic crystal, we modulate $\Delta\phi$ between a certain value and $\pi/2$ at about 2 KHz. Therefore, the current density is modulated between a certain value and zero. By using the lock-in amplifier that is referenced to the phase-modulation frequency, only SH signals that are dependent on $\Delta\phi$ will be detected.

## 2   Effects of the injected carrier and spin densities

In our experiments, the pure spin current is carried by high density carriers injected by the 1500-nm and 750-nm pulses. One natural question is as follows: Can the SH signal be induced simply by the presence of carriers, instead of the spin current? In addition, although both pulses are linearly polarized, and therefore inject spin-unpolarized carriers, the spin transport is known to induce spin accumulation across the laser spot. Therefore, can the SH signal be induced by the spin density, instead of the spin current? In this section, we present our investigations to these questions.

Ideally, the SH signal induced by the carrier density, if it exists, should not be detected by the lock-in amplifier since the carrier density is not modulated. In practice, however, residual modulations of the intensity of the 750-nm pulse exist, due to imperfections of the electro-optic phase modulator. We have independently determined that the residual intensity modulation is on the order of $10^{-5}$.

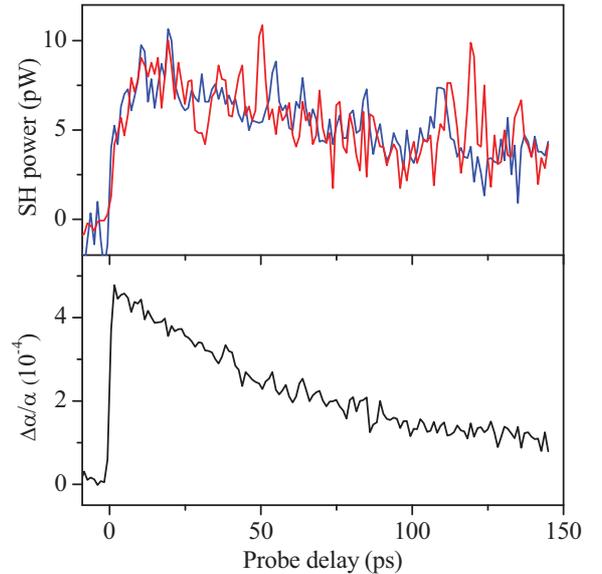

Figure s2: Second-harmonic signal induced by carrier and spin densities.

We investigate SH generation induced by the carrier density by detecting the SH signal with the same setup that is used for the spin-current experiments; the only changes are that the intensity, instead of the phase, of the 750-nm pulse is modulated (with an optical chopper), and the 1500-nm pulse is blocked. The 750-nm pulse is linearly polarized, and injects a peak carrier density of $3.6 \times 10^{17} \text{cm}^{-3}$. A SH signal is indeed observed. The blue curve in the upper panel of Fig. s2 shows the detected SH power as a function of the time delay between the probe and the 750-nm pulses.

This measurement provides both quantitative and qualitative evidences that the SH generation discussed in the main text is not due to the carrier density injected. Quantitatively, the SH power detected with the chopper modulation is less than 10 pW. Since the residual intensity modulation of the electro-optic phase modulator is on the order of $10^{-5}$, the leakage of such a signal in the spin-current experiments should be on the order of 0.1 fW, much smaller than the SH power detected (4 pW, black squares in Fig. 3). Qualitatively, the temporal behaviours of the two SH signals are significantly different. The SH signal measured with the chopper persists for over 100 ps, as expected from the slow carrier recombination, while the SH signal induced by the pure spin current only exists for less than 1 ps, which is consistent with the rapid current relaxation. We have also studied the carrier recombination process by measuring the relative absorption change of the same probe pulse due to the free-carrier absorption, $\Delta\alpha/\alpha$, which is proportional to the carrier density. The result is shown in the lower panel of Fig. s2. Compared to the carrier density, the SH power detected with the chopper modulation has a slower rise and a slower decay. Although the mechanism of the SH generation induced by the carrier density is unclear to us, and is beyond the scope of this study, these features are consistent with modifications of the surface SH generation by carriers trapped at the surface of the sample.

Potential SH generation due to a spin density is also investigated. The red curve in the upper panel of Fig. s2 shows the SH power detected with the chopper modulation and when the 750-nm pulse is circularly polarized. In this case, a same carrier density is injected, but with a spin-polarization of 0.5 due to the well-known spin-selection rule. We observed no difference in SH power between the linear and circular 750-nm pulses, proving that the spin density induces a SH power much smaller than that induced by the carrier density.

## 3 Effect of a transverse charge current

A pure spin current along $\hat{x}$ generates a transverse charge current along $\hat{y}$ due to an inverse spin Hall effect. In addition, the quantum interference of the two current-injecting pulses also contributes to the transverse charge current. It has been well established, both theoretically[23] and experimentally[24], that the density of the transverse charge current is at least one order of magnitude smaller than the pure spin current. Nevertheless, it is still important to exclude such an effect.

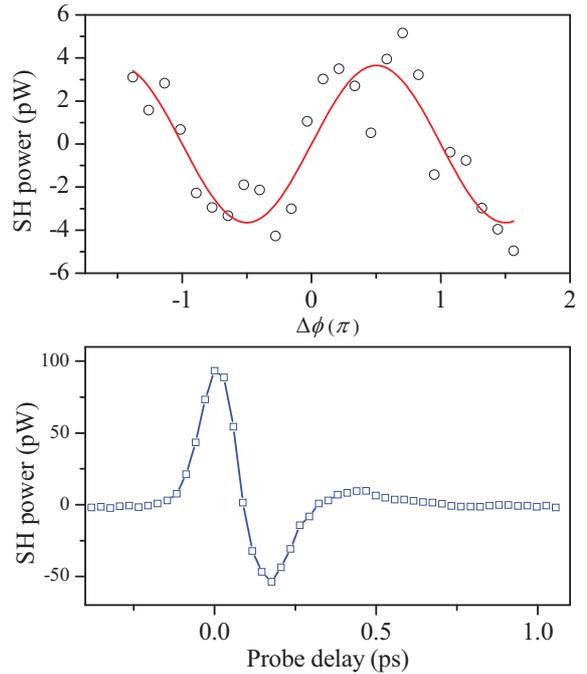

To investigate the behaviour of such a charge current, we intentionally inject a charge current along $\hat{y}$ under the same conditions, except the polarization of the 1500-nm pulse is rotated from $\hat{x}$ to $\hat{y}$[25,26]. In this configuration, instead of injecting a pure spin current along $\hat{x}$, a pure charge current is injected along $\hat{y}$. The average velocity of electrons in this configuration is that of the pure spin current injected with the $\hat{x}$-polarized 1500-nm pulses[25,26]. Therefore, the density of this charge current is comparable to the current density of each spin system in the pure spin current injected with the $\hat{x}$-polarized 1500-nm pulse, and is therefore at least one order of magnitude higher than the transverse charge current accompanying the pure spin current[23,24].

We detected a SH signal induced by such an intentionally injected charge current along $\hat{y}$, under the same experimental conditions as in the spin-current experiments. The upper panel of Fig. s3 shows the detected SH power as a function of $\Delta\phi$. In this measurement, the probe delay is 0.05 ps, which is the delay corresponding to the maximum signal, and the current-injecting and probe spots are overlapped. The magnitude of the SH signal is 4 pW, about three times smaller than the SH power observed in the spin-current experiments under the same conditions (down-triangles in Fig. 3, with a carrier density of $7.2 \times 10^{17} \text{cm}^{-3}$). Since this "simulated" transverse charge current is at least one order of magnitude larger than the actual transverse charge current accompanying the pure spin current in our spin-current experiments, it is safe to conclude that the SH power induced by the latter charge current should be at least one order of magnitude lower than the detected signal in the spin-current experiments.

Figure s3: Second-harmonic signal induced by a transverse charge current.

In order to find additional evidence to distinguish the two contributions, we also investigated the temporal behaviour of the SH signal induced by the transverse charge current. The signal shown in the upper panel of Fig. s3 is too weak for such a time-resolved study. Since the charge current is along $\hat{y}$, we change the polarization of the probe pulse from $\hat{x}$ to $\hat{y}$, so that it is parallel to the current. The SH power is significantly increased. The lower panel of Fig. s3 shows the SH power as a function of probe delay. In this measurement we keep $\Delta\phi = \pi/2$ so that a maximum charge current is injected. Clearly, the temporal behaviour is significantly different from the SH signal observed in the spin-current experiments.

We attribute the observed SH signal associated with the charge current to the well-known electric-field-induced SH generation process[27]. The laser pulses excite electrons to the conduction band with an average velocity along $\hat{y}$, leaving holes in the valence band with opposite momentum. Once injected, electrons and holes move along opposite directions. The space charge field induced by charge separation causes SH generation. Therefore, the temporal behaviour of the SH power reflects the dynamics of the space charge field. In such a plasma oscillation, the space charge field is expected to oscillate, as observed in Fig. s3. Detailed analysis of such a plasma oscillation is beyond the scope of this document. However, the fundamentally different temporal behavious prove that the SH signals discussed in the main text are not induced by the accompanying transverse charge current.

## Supplementary Notes